\documentclass[ALICE,manyauthors]{cernphprep}
\usepackage[comma,square,numbers,sort&compress]{natbib}
\usepackage{hyperref}
\usepackage{lineno}
\usepackage{xspace}
\usepackage{comment}
\usepackage{multirow}
\usepackage{array,booktabs}

\usepackage{amsmath}
\usepackage{bm}

\usepackage{orcidlink}

\graphicspath{ {./figures/} }

\makeatletter
\g@addto@macro\bfseries\boldmath
\makeatother

\usepackage{xcolor}
\definecolor{myOrange}{rgb}{1,0.5,0.}
\definecolor{myGreen}{rgb}{0.0,0.6,0.1}

\newcommand{\removetext}[1]{} 

\newcommand{\pp}           {\ensuremath{\mathrm{pp}}\xspace}

\newcommand{\PbPb}         {\mbox{Pb--Pb}\xspace}

\newcommand{\aaa}{\mbox{A--A}\xspace}

\newcommand{\nineH}        {$\sqrt{s}~=~0.9$~Te\kern-.1emV\xspace}
\newcommand{\seven}        {$\sqrt{s}~=~7$~Te\kern-.1emV\xspace}
\newcommand{\twoH}         {$\sqrt{s}~=~0.2$~Te\kern-.1emV\xspace}
\newcommand{\twosevensix}  {$\sqrt{s}~=~2.76$~Te\kern-.1emV\xspace}
\newcommand{\five}         {$\sqrt{s}~=~5.02$~Te\kern-.1emV\xspace}
\newcommand{\twosevensixnn}{$\sqrt{s_{\mathrm{NN}}}~=~2.76$~Te\kern-.1emV\xspace}
\newcommand{\fivenn}       {$\sqrt{s_{\mathrm{NN}}}~=~5.02$~Te\kern-.1emV\xspace}

\newcommand{\MeVc}         {Me\kern-.1emV/$c$\xspace}
\newcommand{\TeV}          {Te\kern-.1emV\xspace}
\newcommand{\GeV}          {Ge\kern-.1emV\xspace}
\newcommand{\MeV}          {Me\kern-.1emV\xspace}
\newcommand{\GeVmass}      {Ge\kern-.2emV/$c^2$\xspace}
\newcommand{\MeVmass}      {Me\kern-.2emV/$c^2$\xspace}

\newcommand {\etajet} {\ensuremath{\eta_{\mathrm{jet}}}}
\newcommand {\Ntrig} {\ensuremath{N_{\mathrm{trig}}}}

\newcommand {\dphi}        {\ensuremath{\Delta\varphi}\xspace}

\newcommand{\sqrtsNN}{\ensuremath{\sqrt{s_\mathrm{NN}}}}
\newcommand{\sqrts}{\ensuremath{\sqrt{s}}}

\newcommand{\pT}{\ensuremath{p_\mathrm{T}}}

\newcommand{\pTjet}{\ensuremath{p_\mathrm{T, jet}}}
\newcommand{\pTjetch}{\ensuremath{p_\mathrm{T, ch~jet}}}

\newcommand{\Ajet}{\ensuremath{A_\mathrm{jet}}}

\newcommand{\rhoA}{\ensuremath{\rho\Ajet}}

\newcommand{\cRef}{\ensuremath{c_\mathrm{Ref}}}

\newcommand{\pTtrig}{\ensuremath{p_\mathrm{T}^\mathrm{trig}}}

\newcommand{\gev}{\ensuremath{\mathrm{GeV/}c}}

\newcommand{\kT}{\ensuremath{k_\mathrm{T}}}
\newcommand{\antikT}{anti-\ensuremath{k_\mathrm{T}}}

\newcommand{\rr}{\ensuremath{R}}

\newcommand{\pTlow}{\ensuremath{p_\mathrm{T,low}}}
\newcommand{\pThigh}{\ensuremath{p_\mathrm{T,high}}}

\newcommand{\dNjetdDphidpT}{\ensuremath{\frac{{\rm d}^{2}N_{\mathrm{jet}}}{ {\mathrm{d}\pTjet} {\mathrm{d}\dphi}}}}

\newcommand{\IAA}{\ensuremath{{I}_{\rm AA}}}

\newcommand{\IAApT}{\ensuremath{{I}_{\rm AA}(\pTjetch)}}
\newcommand{\IAAphi}{\ensuremath{{I}_{\rm AA}(\dphi)}}

\newcommand{\qhat}{\ensuremath{\hat{q}}}

\newcommand{\Drecoil}{\ensuremath{\Delta_\mathrm{recoil}}}

\newcommand{\DrecoilpTphi}{\ensuremath{\Delta_\mathrm{recoil}(\pTjet,\dphi)}}

\newcommand{\Drecoilphi}{\ensuremath{\Delta_\mathrm{recoil}(\dphi)}}

\newcommand{\DrecoilpTchphi}{\ensuremath{\Delta_\mathrm{recoil}(\pTjetch,\dphi)}}
\newcommand{\DrecoilpTch}{\ensuremath{\Delta_\mathrm{recoil}(\pTjetch)}}

\newcommand{\TTSig}{\ensuremath{\mathrm{TT}_\mathrm{sig}}}
\newcommand{\TTRef}{\ensuremath{\mathrm{TT}_\mathrm{ref}}}

\newcommand{\TT}[2]{\ensuremath{\mathrm{TT}\{#1,#2\}}}

\newcommand{\pTjetin}[2]{\ensuremath{#1<\pTjetch<#2}}

\begin{document}

\begin{titlepage}
\PHyear{2023}       
\PHnumber{189}      
\PHdate{29 August}  

\title{Observation of medium-induced yield enhancement and acoplanarity broadening of low-$\mathbf{p_\mathrm{T}}$ jets from measurements in pp and central \PbPb\ collisions at $\mathbf{\sqrt{\textit{s}_{\mathrm{\mathbf{NN}}}}} \mathbf{\;= 5.02}$ TeV}
\ShortTitle{Jet yield enhancement and acoplanarity broadening}  
\Collaboration{ALICE Collaboration\thanks{See Appendix~\ref{app:collab} for the list of collaboration members}}
\ShortAuthor{ALICE Collaboration} 

\begin{abstract}

The ALICE Collaboration reports the measurement of semi-inclusive distributions of charged-particle jets recoiling from a high transverse momentum (high \pT) hadron trigger in proton--proton and central \PbPb\ collisions at $\sqrtsNN=5.02$ TeV. A data-driven statistical method is used to mitigate the large uncorrelated background in central \PbPb\ collisions. Recoil jet distributions are reported for jet resolution parameter $\rr=0.2$, 0.4, and 0.5 in the range $7<\pTjet<140$ \gev\ and trigger--recoil jet azimuthal separation $\pi/2<\dphi<\pi$. The measurements exhibit a marked medium-induced jet yield enhancement at low \pT\ and at large azimuthal deviation from $\dphi\sim\pi$. The enhancement is characterized by its dependence on \dphi, which has a slope that differs from zero by 4.7$\sigma$. Comparisons to model calculations incorporating different formulations of jet quenching are reported. These comparisons indicate that the observed yield enhancement arises from the response of the QGP medium to jet propagation.

\end{abstract}

\end{titlepage}

\setcounter{page}{2} 


Matter at very high temperature forms a quark--gluon plasma (QGP), the state of matter in which quarks and gluons are not bound in colorless hadrons~\cite{Busza:2018rrf,Harris:2023tti}. A QGP filled the early Universe a few microseconds after the Big Bang, and is generated today in high-energy nuclear collisions at the Large Hadron Collider (LHC) and the Relativistic Heavy Ion Collider (RHIC)~\cite{BRAHMS:2004adc,PHENIX:2004vcz,PHOBOS:2004zne,STAR:2005gfr,ALICE:2022wpn}. Measurements at RHIC and the LHC and their comparison to theoretical calculations show that the QGP flows with low specific shear viscosity~\cite{Heinz:2013th}. Quantum chromodynamics (QCD) calculations on the lattice show that the effective number of QGP degrees of freedom is $\sim$15\% lower than that of freely-interacting quarks and gluons, at temperatures well above the deconfinement transition temperature $\sim150$ MeV~\cite{Borsanyi:2013bia,HotQCD:2014kol}. However, understanding the origin of such emergent phenomena in terms of quasi-particle degrees of freedom remains elusive.

QCD jets arise from hard (high momentum-transfer $Q^2$) scattering of quarks and gluons (partons). The highly-virtual scattered partons radiate a gluon shower that hadronizes into a correlated spray of experimentally observable hadrons. Jet measurements in proton--proton (\pp) collisions provide stringent tests of perturbative QCD (pQCD) calculations~\cite{CMS:2016jip,ATLAS:2017ble,Acharya:2019jyg}.
In nucleus--nucleus (\aaa) collisions jets interact with the QGP, generating observable modifications to jet production and structure (``jet quenching'')~\cite{Cunqueiro:2021wls}. Comparison of jet quenching measurements and calculations provides unique insight into QGP dynamics and transport properties~\cite{Majumder:2010qh,JETSCAPE:2021ehl}.

Measurements of medium-induced jet angular deflection and substructure modification may elucidate microscopic QGP structure ~\cite{Appel:1985dq,Blaizot:1986ma,DEramo:2018eoy}. Jet scattering off of QGP quasi-particles is the partonic analog to Rutherford scattering off of atomic nuclei ~\cite{Rutherford:1911zz}. However, such measurements are challenging in heavy-ion collisions, due to large uncorrelated background. This is especially the case for jets with low transverse momentum (\pTjet), for which deflection effects may be sizable.

In this Letter the ALICE Collaboration reports measurements of the semi-inclusive distribution of charged-particle jets recoiling from a high-\pT\ hadron trigger~\cite{Adam:2015doa, Adamczyk:2017yhe} in inelastic \pp\ and in central \PbPb\ collisions at center-of-mass energy per nucleon--nucleon collision $\sqrtsNN=5.02$ TeV. Uncorrelated jet yield in central \PbPb\ collisions is corrected using a statistical approach~\cite{Adamczyk:2017yhe}, which enables precise recoil jet measurements at low \pTjet\ and large jet radius \rr, allowing for a comprehensive search for jet deflection effects over broad phase space. 

Recoil jet yield distributions are measured as a function of \pTjet\ and acoplanarity \dphi, the azimuthal separation of the trigger hadron and recoil jet, for jet resolution parameters $\rr=0.2$, 0.4, and 0.5. Recoil jet measurements are reported as a function of \pTjet\ for $7<\pTjet<140$ \gev\ within $|\dphi - \pi| < 0.6$ and as a function of \dphi\ for $\pi/2<\dphi<\pi$ within $10<\pTjet<100$ \gev. Theoretical  calculations incorporating jet quenching are compared to the data. Analysis details and  additional physics results are reported in a companion article~\cite{LonghJetpaper}.

The ALICE apparatus and its performance are described in Refs.~\cite{ALICE:2008ngc,ALICE:2014sbx}. The data for \pp\ collisions at $\sqrts=5.02$ TeV were recorded during the 2015 and 2017 LHC runs using a minimum bias (MB) trigger~\cite{LonghJetpaper}. The data for \PbPb\ collisions at $\sqrtsNN=5.02$ TeV were recorded during the 2018 run using MB and centrality-enhanced triggers~\cite{LonghJetpaper}. The \PbPb\ event population is selected for high event activity in the forward V0 detectors, corresponding to the 10\% most-central fraction of the total \PbPb\ hadronic interaction cross section. After offline event selection, the analyzed dataset has 1.04B events for \pp\ collisions and 89M events for central \PbPb\ collisions.

Charged-particle tracks are reconstructed from hits in the ALICE Inner Tracking System (ITS) and Time Projection Chamber (TPC). The response of these detectors was non-uniform in azimuth and varied between data-taking runs. Tracks are selected to account for such variations, resulting in uniform and stable tracking efficiency~\cite{LonghJetpaper}. Tracks are accepted within pseudorapidity $|\eta|<0.9$ and $\pT>0.15$ \gev.

The same analysis is carried out on \pp\ and central \PbPb\ events.
Events are selected based on the presence of a high-\pT\ charged-hadron trigger track within $\pTlow<\pT<\pThigh$, denoted \TT{\pTlow}{\pThigh} (``Trigger Track,'' units in \gev). For events with multiple such tracks, one track is chosen randomly as the trigger. The \pT-dependence of the resulting TT distribution corresponds to that of inclusive charged-particle production. The analysis utilizes two TT classes, \TT{20}{50}, denoted ``signal'', and \TT{5}{7}, denoted ``reference.''

For TT-selected events, jet reconstruction with charged tracks is carried out in two passes, using the \kT\ and \antikT\ jet reconstruction algorithms and the \pT\ recombination scheme~\cite{FastJet,Cacciari:2008gp,Cacciari:2011ma}. The jet acceptance is $|\etajet|<0.9-\rr$ over the full azimuth, with additional selection on jet area to suppress unphysical jets~\cite{Adam:2015doa}. Jets containing tracks with $\pT>100$ \gev\ are rejected; this rejection has negligible effect on the reported results. There is no other rejection of individual jet candidates.

The first reconstruction pass utilizes the \kT\ algorithm to estimate the event-wise median \pT\ density $\rho$~\cite{Cacciari:2007fd,Adam:2015doa}. The signal and reference TT-selected event populations have different hard jet distributions, which influence the $\rho$ distribution~\cite{Adamczyk:2017yhe,LonghJetpaper}. Precise correction for uncorrelated background yield in central \PbPb\ collisions requires a shift in the reference-TT $\rho$ distribution, determined by a data-driven procedure with sub-per mil precision~\cite{LonghJetpaper}. This effect is negligible in \pp\ collisions. The second reconstruction pass generates the jet population for physics analysis, utilizing the \antikT\ algorithm with $\rr=0.2$, 0.4, and 0.5. The \pT\ of each second-pass jet is adjusted by a rough estimate of the background contribution \rhoA, where $A$ is the jet area. This estimate is refined by unfolding, discussed below.

Recoil jet distributions are normalized by the corresponding number of triggers and are semi-inclusive; absent of background they correspond to the production cross-section ratio for hadron--jet coincidences and inclusive hadrons~\cite{Adam:2015doa} and are perturbatively calculable. The observable \Drecoil\ is defined as the difference of such signal-TT and reference-TT distributions~\cite{Adam:2015doa}:

\begin{equation}
\DrecoilpTphi = 
\frac{1}{\Ntrig}\dNjetdDphidpT\Bigg\vert_{\pTtrig\in{\TTSig}} -
\cRef\times \frac{1}{\Ntrig}\dNjetdDphidpT\Bigg\vert_{\pTtrig\in{\TTRef}}.
\label{eq:DRecoil}
\end{equation}

\noindent
The scale factor \cRef\ is extracted from data following the data-driven procedure described in Refs.~\cite{Adam:2015doa,LonghJetpaper}.
After scaling by \cRef, the distribution of background jet yield that is uncorrelated with the trigger is identical in the two terms. The subtraction in \Drecoil\ therefore provides precise correction for this background yield, enabling recoil jet measurements at low \pTjet\ and large \rr.

Multiple hard partonic interactions (MPIs) in the same nuclear collision are independent and do not interfere~\cite{Collins:1989gx}. MPIs which generate an uncorrelated trigger hadron and recoil jet in the same event constitute a significant background in the search for large-angle jet deflection, since the MPI-generated \dphi\ distribution is uniform, masking any \dphi-dependent physical effect. However, \Drecoil\ corrects the yield due to all uncorrelated sources, including MPIs, and no additional correction procedure to account for the MPI contribution is warranted in the analysis. 

The measured \Drecoil\ distribution is smeared in \pTjet\ and \dphi\ due to detector effects and residual background fluctuations~\cite{Adam:2015doa,Adamczyk:2017yhe}. Correction for this smearing is carried out using iterative Bayesian unfolding~\cite{Adye:2011gm} in one dimension (\pTjetch) for measuring \DrecoilpTch, and in two dimensions (\pTjetch,\dphi) for measuring \Drecoilphi; see Ref.~\cite{LonghJetpaper} for details and consistency checks. The largest systematic uncertainty in the corrected \Drecoil\ distribution for \pp\ collisions is due to tracking efficiency, while that for \PbPb\ collisions is due to the choice of prior used for unfolding. 

The measurements are compared to theoretical model calculations incorporating jet quenching. All models  generate hard processes using PYTHIA8 (Monash tune~\cite{Sjostrand:2014zea,Skands:2014pea}), but differ in the treatment of jet--medium interactions and QGP medium response. JEWEL~\cite{Zapp:2008gi,Zapp:2013vla} calculates in-medium scattering using pQCD matrix elements. JETSCAPE~\cite{JETSCAPE:2021ehl} incorporates a virtuality-dependent interaction  based on MATTER~\cite{Majumder:2013re,Majumder:2009zu} and LBT~\cite{He:2015pra,Wang:2013cia}. The Hybrid Model~\cite{Casalderrey-Solana:2014bpa} describes weakly-coupled jet dynamics perturbatively, with strongly-coupled jet--medium interactions based on the AdS/CFT correspondence. JEWEL calculations optionally include medium response (``recoils on'' or ``recoils off''), where the ``recoils on'' calculation follows the ``4MomSub'' prescription~\cite{KunnawalkamElayavalli:2017hxo}. The Hybrid Model likewise optionally includes medium response (``wake'') and elastic scattering from discrete scattering centers~\cite{DEramo:2018eoy}. Comparison is also made to a leading-order (LO) pQCD calculation with Sudakov resummation, in which medium-induced broadening is controlled by the jet transport coefficient \qhat~\cite{Chen:2016vem}.

\begin{figure*}[htb]
\begin{center}
\includegraphics[width=0.95\textwidth]{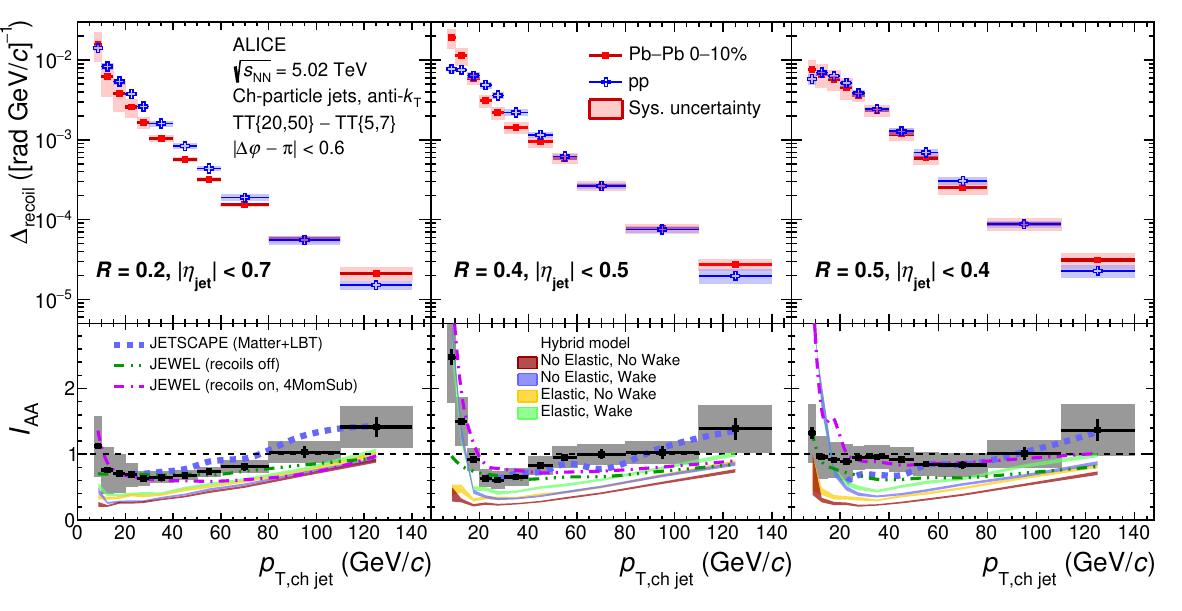}
 \end{center}
\caption{Distributions of recoil jets with $\rr=0.2$, 0.4, and 0.5 in \pp\ and central \PbPb\ collisions at $\sqrtsNN=5.02$ TeV.
Upper panels: corrected \DrecoilpTch\ distributions. Lower panels: \IAApT\ (see text). Also shown are calculations based on JETSCAPE~\cite{JETSCAPE:2021ehl}, JEWEL~\cite{Zapp:2008gi,Zapp:2013vla}, and the Hybrid model~\cite{Casalderrey-Solana:2014bpa} .}
\label{fig:IAApT}
\end{figure*}

Figure~\ref{fig:IAApT}, upper panels, show \DrecoilpTch, the \DrecoilpTchphi\ distribution integrated over $|\dphi-\pi|<0.6$, for $\rr=0.2$, 0.4, and 0.5 in \pp\ and central \PbPb\ collisions at $\sqrtsNN=5.02$ TeV. The distributions cover $7<\pTjetch<140$ \gev, including the lowest reported \pTjet\ value for jet measurements in heavy-ion collisions at the LHC. The distributions are qualitatively similar, though with shape differences for $\pTjetch\lesssim30$ \gev.

Figure~\ref{fig:IAApT}, lower panels,  show \IAApT, the ratio of the \PbPb\ and \pp\ \DrecoilpTch\ distributions. In the range $\pTjetch<20$ \gev, \IAA\ is consistent with or above unity for all \rr. For $20<\pTjetch\lesssim60$ \gev, \IAA\ is below unity for $\rr=0.2$ and 0.4, which is usually interpreted as medium-induced yield suppression due to energy loss~\cite{Adam:2015doa}. The value of \IAA\ is consistent with or above unity at higher \pTjetch\ for $\rr=0.2$ and 0.4, and at all \pTjetch\ for $\rr=0.5$. It is shown in Ref.~\cite{He:2024rcv} that energy loss of the trigger-side jet can enhance \IAA\, and it is expected that jets with \IAA\ equal to or even above unity may still experience energy loss, consistent with inclusive jet measurements. It also suggests that increasing \IAApT\ with increasing \pTjetch\ may indicate evolution in the geometric (``surface'') bias of vertices which generate the observed high-\pT\ hadron triggers~\cite{LonghJetpaper}.
The \IAApT\ distributions for $\rr=0.2$ and 0.4 exhibit broad minima near $\pTjetch\sim$ 20--30 \gev; comparisons with models above and below this minimum are discussed separately. 

In the range $\pTjetch>20$ \gev, for $\rr=0.2$ and 0.4 JETSCAPE and the Hybrid Model (all options) exhibit a similar increase in \IAApT\ with increasing \pTjetch\ as the data. JETSCAPE also reproduces the magnitude of \IAApT, while the Hybrid Model predicts a smaller value. JEWEL (recoils off) agrees with the measured \IAApT\ up to 80 \gev\ for $\rr=0.2$ and up to 40 \gev\ for $\rr=0.4$, but underpredicts it at higher \pTjetch. JEWEL (recoils on) similarly underpredicts the data in $\pTjetch>50$ \gev. For $\rr=0.5$, JETSCAPE describes the data in $\pTjetch>50$ \gev, but underpredicts it below that range. JEWEL (recoils on) accurately describes the measured \IAA\ in $\pTjetch>20$ \gev\ for $\rr=0.5$, while JEWEL (recoils off) underpredicts it.

For $\pTjetch<20$ \gev, the data exhibit an increase in \IAApT\ with decreasing \pTjetch\ for $\rr=0.4$, with a less significant or negligible increase for $\rr=0.2$ and 0.5. However, the difference in the magnitude of \IAApT\ between different $R$ jets is not significant within uncertainties. Notably, the Hybrid Model with wake-on (both with and without elastic scattering) and JEWEL (recoils on) reproduce the data for $\rr=0.4$. This suggests that the increase in \IAApT\ towards low \pTjetch\ may arise from medium response to interactions of higher-energy jets that are correlated with the trigger, although these models do less well at reproducing the low-\pTjetch\ \IAApT\ for $\rr=0.5$ jets, indicating that the redistribution of jet energy is not fully captured by models.

\begin{figure*}[htb]
\begin{center}
\includegraphics[width=0.95\textwidth]{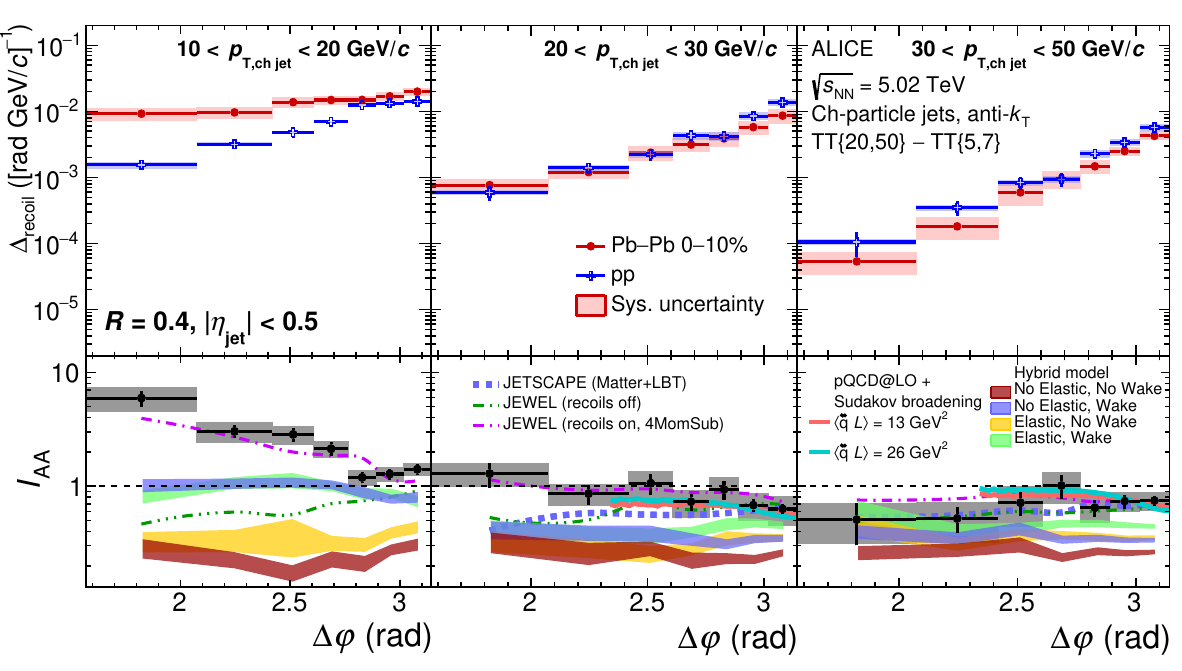}
\end{center}
\caption{Upper panels: Corrected \Drecoilphi\ distributions for $\rr=0.4$ in \PbPb\ and \pp\ collisions at $\sqrtsNN=5.02$ TeV, for intervals in recoil \pTjetch:
[10,20] (left), [20,30] (middle), and [30,50] (right) \gev. Lower panels: \IAAphi. Predictions from JETSCAPE~\cite{JETSCAPE:2021ehl}, JEWEL~\cite{Zapp:2008gi,Zapp:2013vla}, and the LO pQCD calculation~\cite{Chen:2016vem} are also shown.}
\label{fig:IAAphi}
\end{figure*}

Figure~\ref{fig:IAAphi}, upper panels, show \Drecoilphi, the \DrecoilpTchphi\ distribution projected onto \dphi\ in intervals of \pTjetch, for $\rr=0.4$ in \pp\ and central \PbPb\ collisions. The lower panels show their ratio, \IAAphi. For $30<\pTjetch<50$ \gev, medium-induced yield suppression ($\IAAphi<1$) is observed, largely independent of \dphi. For $20<\pTjetch<30$ \gev, suppression is observed at $\dphi\sim\pi$, with a gradual but significant increase of \IAAphi\ at larger deviation from $\dphi\sim\pi$. Notably, for $10<\pTjetch<20$ \gev, a marked medium-induced excess is observed ($\IAAphi>1$), which increases with increasing deviation from $\dphi\sim\pi$. A linear fit of this distribution in the range $0.5 \pi<\dphi<0.92\pi$, taking into account uncorrelated uncertainties only, has slope $-40.5{\pm}8.6$, differing by 4.7$\sigma$ from zero (which corresponds to no medium-induced modification). This is the first observation of strong acoplanarity broadening in the QGP. 

The data in Fig.~\ref{fig:IAApT}, middle panels, and in Fig.~\ref{fig:IAAphi} are slices of the same two-dimensional distributions \DrecoilpTphi. Note that \IAApT\ is integrated over $|\dphi-\pi|<0.6$, corresponding approximately to the rightmost four points in Fig.~\ref{fig:IAAphi}, which should be considered when comparing the figures.

Figure~\ref{fig:IAAphi}, lower panels, also show theoretical calculations. The LO pQCD calculation is consistent with data in \pTjetin{20}{50}~\gev\ and $2.4<\dphi<\pi$ for $13<\langle\qhat{L}\rangle<26$ GeV$^2$, where $L$ is the in-medium path length. 
JETSCAPE overpredicts the suppression in \pTjetin{20}{30} \gev, but agrees with data in \pTjetin{30}{50} \gev. JEWEL (recoils on) describes both the data shape and magnitude well for all \pTjetch\ intervals, including the significant broadening in \pTjetin{10}{20} \gev\ that is not predicted by JEWEL (recoils off). None of the Hybrid Model variants describes the observed broadening at low \pTjetch. These variants generate different magnitude of suppression but underestimate the measured value of \IAA\ in all \pTjetch\ bins. Only JEWEL (recoils on) correctly reproduces the marked azimuthal broadening at low \pTjetch\ seen in data. 

\begin{figure*}[htb]
\begin{center}
\includegraphics[width = 0.9\textwidth]{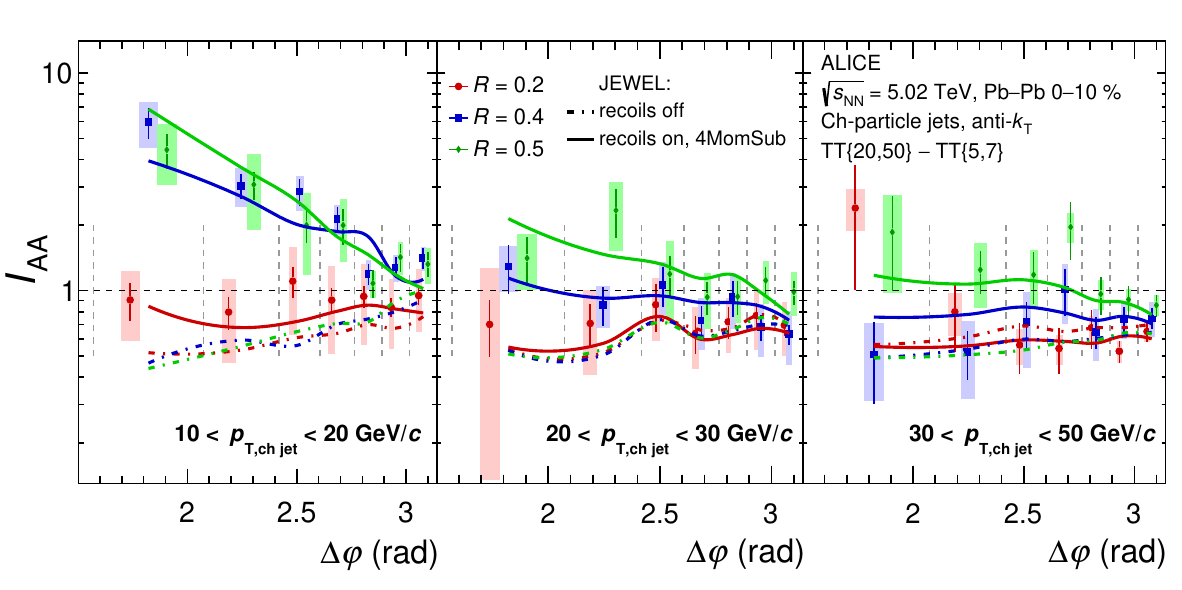}
\end{center}
\caption{\IAAphi\ for $\rr=0.2$, 0.4 and 0.5, for intervals in recoil \pTjetch:
[10,20], [20,30], and [30,50] \gev. The central points and systematic uncertainties are offset from the center of the \dphi\ intervals for clarity. The vertical dashed gray lines represent the \dphi\ interval edges. Predictions from JEWEL are also shown. }
\label{fig:IAAphi_vs_R}
\end{figure*}

Figure~\ref{fig:IAAphi_vs_R} shows \IAAphi\ for $\rr=0.2$, 0.4, and 0.5, for the \pTjetch\ intervals in Fig.~\ref{fig:IAAphi}. The medium-induced acoplanarity broadening in Fig.~\ref{fig:IAAphi}, left panel, is seen only in the range \pTjetin{10}{20} \gev, and only for $\rr=0.4$ and 0.5 . The value of \IAAphi\ is either consistent with unity or suppressed at larger \pTjetch\ for $\rr=0.4$ and 0.5, and for all measured \pTjetch\ for $\rr=0.2$. The JEWEL (recoils on) calculation is likewise consistent within uncertainties with all of these data. 

Figures~\ref{fig:IAApT},~\ref{fig:IAAphi}, and~\ref{fig:IAAphi_vs_R} present the first observation of medium-induced jet yield excess and acoplanarity broadening in the QGP.
The broadening is significant in $10<\pTjetch<20$ \gev\ for $\rr=0.4$ and 0.5  but is negligible for $\rr=0.2$, and at larger \pTjetch\ for all \rr. This rapid transition in the acoplanarity distribution shape as a function \pTjetch\ and \rr\ is striking. Possible medium-induced acoplanarity broadening mechanisms include jet scattering from QGP quasi-particles; wake effects~\cite{Cao:2020wlm}; and jet splitting.

The latter two mechanisms do not generate perturbatively interpretable jets, with constituents that are softer in \pT\ and spatially more diffuse. In these scenarios, the rate to generate a correlated ``jet'' with $\pTjetch>10$~\gev\ may scale approximately with the jet area, i.e. $\rr^2$, resulting in a strong \rr-dependence of the \IAAphi\ enhancement at low \pTjetch, as observed. In contrast, a strong \rr-dependence of the \IAAphi\ enhancement is not a natural consequence of jet scattering from QGP quasi-particles, which should generate similar effects for $\rr=0.2$, 0.4, and 0.5. The observed systematic dependence therefore disfavors in-medium jet scattering as the primary origin of in-medium acoplanarity broadening. Both JEWEL and the Hybrid model describe the observed low-\pTjetch\ behavior of \IAApT, only if jet--medium response is included. None of the models considered here successfully describe all available data.

In summary, measurements of semi-inclusive distributions of charged-particle jets recoiling from a high-\pT\ hadron trigger in \pp\ and central \PbPb\ collisions at $\sqrtsNN=5.02$ TeV have been reported over a broad kinematic range, including low \pTjet\ and large \rr.  A marked medium-induced enhancement in recoil jet acoplanarity is observed for the first time, but only at low \pTjet\ for large \rr; this favors QGP wake effects or jet splitting as the underlying physical mechanism, and disfavors large-angle jet scattering. 

Current model calculations incorporating jet quenching do not reproduce all of these observations. Further modeling developments, and their comparison to these and similar data, promise significant new understanding of the mechanisms governing energy transport and the dynamics of the QGP.


\newenvironment{acknowledgement}{\relax}{\relax}
\begin{acknowledgement}
\section*{Acknowledgements}

We thank 
Daniel Pablos, 
Krishna Rajagopal, 
Zachary Hulcher, 
Shuyi Wei, 
and Guangyou Qin 
for providing theoretical calculations.
We thank the JETSCAPE collaboration for guidance in using the JETSCAPE framework, and Raghav Kunnawalkam Elayavalli for providing the medium parameters for JEWEL simulations.


The ALICE Collaboration would like to thank all its engineers and technicians for their invaluable contributions to the construction of the experiment and the CERN accelerator teams for the outstanding performance of the LHC complex.
The ALICE Collaboration gratefully acknowledges the resources and support provided by all Grid centres and the Worldwide LHC Computing Grid (WLCG) collaboration.
The ALICE Collaboration acknowledges the following funding agencies for their support in building and running the ALICE detector:
A. I. Alikhanyan National Science Laboratory (Yerevan Physics Institute) Foundation (ANSL), State Committee of Science and World Federation of Scientists (WFS), Armenia;
Austrian Academy of Sciences, Austrian Science Fund (FWF): [M 2467-N36] and Nationalstiftung f\"{u}r Forschung, Technologie und Entwicklung, Austria;
Ministry of Communications and High Technologies, National Nuclear Research Center, Azerbaijan;
Conselho Nacional de Desenvolvimento Cient\'{\i}fico e Tecnol\'{o}gico (CNPq), Financiadora de Estudos e Projetos (Finep), Funda\c{c}\~{a}o de Amparo \`{a} Pesquisa do Estado de S\~{a}o Paulo (FAPESP) and Universidade Federal do Rio Grande do Sul (UFRGS), Brazil;
Bulgarian Ministry of Education and Science, within the National Roadmap for Research Infrastructures 2020--2027 (object CERN), Bulgaria;
Ministry of Education of China (MOEC) , Ministry of Science \& Technology of China (MSTC) and National Natural Science Foundation of China (NSFC), China;
Ministry of Science and Education and Croatian Science Foundation, Croatia;
Centro de Aplicaciones Tecnol\'{o}gicas y Desarrollo Nuclear (CEADEN), Cubaenerg\'{\i}a, Cuba;
Ministry of Education, Youth and Sports of the Czech Republic, Czech Republic;
The Danish Council for Independent Research | Natural Sciences, the VILLUM FONDEN and Danish National Research Foundation (DNRF), Denmark;
Helsinki Institute of Physics (HIP), Finland;
Commissariat \`{a} l'Energie Atomique (CEA) and Institut National de Physique Nucl\'{e}aire et de Physique des Particules (IN2P3) and Centre National de la Recherche Scientifique (CNRS), France;
Bundesministerium f\"{u}r Bildung und Forschung (BMBF) and GSI Helmholtzzentrum f\"{u}r Schwerionenforschung GmbH, Germany;
General Secretariat for Research and Technology, Ministry of Education, Research and Religions, Greece;
National Research, Development and Innovation Office, Hungary;
Department of Atomic Energy Government of India (DAE), Department of Science and Technology, Government of India (DST), University Grants Commission, Government of India (UGC) and Council of Scientific and Industrial Research (CSIR), India;
National Research and Innovation Agency - BRIN, Indonesia;
Istituto Nazionale di Fisica Nucleare (INFN), Italy;
Japanese Ministry of Education, Culture, Sports, Science and Technology (MEXT) and Japan Society for the Promotion of Science (JSPS) KAKENHI, Japan;
Consejo Nacional de Ciencia (CONACYT) y Tecnolog\'{i}a, through Fondo de Cooperaci\'{o}n Internacional en Ciencia y Tecnolog\'{i}a (FONCICYT) and Direcci\'{o}n General de Asuntos del Personal Academico (DGAPA), Mexico;
Nederlandse Organisatie voor Wetenschappelijk Onderzoek (NWO), Netherlands;
The Research Council of Norway, Norway;
Commission on Science and Technology for Sustainable Development in the South (COMSATS), Pakistan;
Pontificia Universidad Cat\'{o}lica del Per\'{u}, Peru;
Ministry of Education and Science, National Science Centre and WUT ID-UB, Poland;
Korea Institute of Science and Technology Information and National Research Foundation of Korea (NRF), Republic of Korea;
Ministry of Education and Scientific Research, Institute of Atomic Physics, Ministry of Research and Innovation and Institute of Atomic Physics and University Politehnica of Bucharest, Romania;
Ministry of Education, Science, Research and Sport of the Slovak Republic, Slovakia;
National Research Foundation of South Africa, South Africa;
Swedish Research Council (VR) and Knut \& Alice Wallenberg Foundation (KAW), Sweden;
European Organization for Nuclear Research, Switzerland;
Suranaree University of Technology (SUT), National Science and Technology Development Agency (NSTDA) and National Science, Research and Innovation Fund (NSRF via PMU-B B05F650021), Thailand;
Turkish Energy, Nuclear and Mineral Research Agency (TENMAK), Turkey;
National Academy of  Sciences of Ukraine, Ukraine;
Science and Technology Facilities Council (STFC), United Kingdom;
National Science Foundation of the United States of America (NSF) and United States Department of Energy, Office of Nuclear Physics (DOE NP), United States of America.
In addition, individual groups or members have received support from:
European Research Council, Strong 2020 - Horizon 2020 (grant nos. 950692, 824093), European Union;
Academy of Finland (Center of Excellence in Quark Matter) (grant nos. 346327, 346328), Finland;
\end{acknowledgement}

\bibliographystyle{utphys}   
\bibliography{Common/bibliography}

\newpage
\appendix

\section{The ALICE Collaboration}
\label{app:collab}
\begin{flushleft} 
\small

S.~Acharya\,\orcidlink{0000-0002-9213-5329}\,$^{\rm 128}$, 
D.~Adamov\'{a}\,\orcidlink{0000-0002-0504-7428}\,$^{\rm 87}$, 
G.~Aglieri Rinella\,\orcidlink{0000-0002-9611-3696}\,$^{\rm 33}$, 
M.~Agnello\,\orcidlink{0000-0002-0760-5075}\,$^{\rm 30}$, 
N.~Agrawal\,\orcidlink{0000-0003-0348-9836}\,$^{\rm 52}$, 
Z.~Ahammed\,\orcidlink{0000-0001-5241-7412}\,$^{\rm 136}$, 
S.~Ahmad\,\orcidlink{0000-0003-0497-5705}\,$^{\rm 16}$, 
S.U.~Ahn\,\orcidlink{0000-0001-8847-489X}\,$^{\rm 72}$, 
I.~Ahuja\,\orcidlink{0000-0002-4417-1392}\,$^{\rm 38}$, 
A.~Akindinov\,\orcidlink{0000-0002-7388-3022}\,$^{\rm 142}$
% ...

\section*{Affiliation Notes}

$^{\rm I}$ Deceased\\
$^{\rm II}$ Also at: Max-Planck-Institut fur Physik, Munich, Germany\\
$^{\rm III}$ Also at: Italian National Agency for New Technologies, Energy and Sustainable Economic Development (ENEA), Bologna, Italy\\
$^{\rm IV}$ Also at: Dipartimento DET del Politecnico di Torino, Turin, Italy\\
$^{\rm V}$ Also at: Yildiz Technical University, Istanbul, T\"{u}rkiye\\
$^{\rm VI}$ Also at: Department of Applied Physics, Aligarh Muslim University, Aligarh, India\\
$^{\rm VII}$ Also at: Institute of Theoretical Physics, University of Wroclaw, Poland\\
$^{\rm VIII}$ Also at: An institution covered by a cooperation agreement with CERN\\

\section*{Collaboration Institutes}

$^{1}$ A.I. Alikhanyan National Science Laboratory (Yerevan Physics Institute) Foundation, Yerevan, Armenia\\
$^{2}$ AGH University of Krakow, Cracow, Poland\\
$^{3}$ Bogolyubov Institute for Theoretical Physics, National Academy of Sciences of Ukraine, Kiev, Ukraine\\
$^{4}$ Bose Institute, Department of Physics  and Centre for Astroparticle Physics and Space Science (CAPSS), Kolkata, India\\
$^{5}$ California Polytechnic State University, San Luis Obispo, California, United States\\
$^{6}$ Central China Normal University, Wuhan, China\\
$^{7}$ Centro de Aplicaciones Tecnol\'{o}gicas y Desarrollo Nuclear (CEADEN), Havana, Cuba\\
$^{8}$ Centro de Investigaci\'{o}n y de Estudios Avanzados (CINVESTAV), Mexico City and M\'{e}rida, Mexico\\
$^{9}$ Chicago State University, Chicago, Illinois, United States\\
$^{10}$ China Institute of Atomic Energy, Beijing, China\\
$^{11}$ China University of Geosciences, Wuhan, China\\
$^{12}$ Chungbuk National University, Cheongju, Republic of Korea\\
$^{13}$ Comenius University Bratislava, Faculty of Mathematics, Physics and Informatics, Bratislava, Slovak Republic\\
$^{14}$ COMSATS University Islamabad, Islamabad, Pakistan\\
$^{15}$ Creighton University, Omaha, Nebraska, United States\\
$^{16}$ Department of Physics, Aligarh Muslim University, Aligarh, India\\
$^{17}$ Department of Physics, Pusan National University, Pusan, Republic of Korea\\
$^{18}$ Department of Physics, Sejong University, Seoul, Republic of Korea\\
$^{19}$ Department of Physics, University of California, Berkeley, California, United States\\
$^{20}$ Department of Physics, University of Oslo, Oslo, Norway\\
$^{21}$ Department of Physics and Technology, University of Bergen, Bergen, Norway\\
$^{22}$ Dipartimento di Fisica, Universit\`{a} di Pavia, Pavia, Italy\\
$^{23}$ Dipartimento di Fisica dell'Universit\`{a} and Sezione INFN, Cagliari, Italy\\
$^{24}$ Dipartimento di Fisica dell'Universit\`{a} and Sezione INFN, Trieste, Italy\\
$^{25}$ Dipartimento di Fisica dell'Universit\`{a} and Sezione INFN, Turin, Italy\\
$^{26}$ Dipartimento di Fisica e Astronomia dell'Universit\`{a} and Sezione INFN, Bologna, Italy\\
$^{27}$ Dipartimento di Fisica e Astronomia dell'Universit\`{a} and Sezione INFN, Catania, Italy\\
$^{28}$ Dipartimento di Fisica e Astronomia dell'Universit\`{a} and Sezione INFN, Padova, Italy\\
$^{29}$ Dipartimento di Fisica `E.R.~Caianiello' dell'Universit\`{a} and Gruppo Collegato INFN, Salerno, Italy\\
$^{30}$ Dipartimento DISAT del Politecnico and Sezione INFN, Turin, Italy\\
$^{31}$ Dipartimento di Scienze MIFT, Universit\`{a} di Messina, Messina, Italy\\
$^{32}$ Dipartimento Interateneo di Fisica `M.~Merlin' and Sezione INFN, Bari, Italy\\
$^{33}$ European Organization for Nuclear Research (CERN), Geneva, Switzerland\\
$^{34}$ Faculty of Electrical Engineering, Mechanical Engineering and Naval Architecture, University of Split, Split, Croatia\\
$^{35}$ Faculty of Engineering and Science, Western Norway University of Applied Sciences, Bergen, Norway\\
$^{36}$ Faculty of Nuclear Sciences and Physical Engineering, Czech Technical University in Prague, Prague, Czech Republic\\
$^{37}$ Faculty of Physics, Sofia University, Sofia, Bulgaria\\
$^{38}$ Faculty of Science, P.J.~\v{S}af\'{a}rik University, Ko\v{s}ice, Slovak Republic\\
$^{39}$ Frankfurt Institute for Advanced Studies, Johann Wolfgang Goethe-Universit\"{a}t Frankfurt, Frankfurt, Germany\\
$^{40}$ Fudan University, Shanghai, China\\
$^{41}$ Gangneung-Wonju National University, Gangneung, Republic of Korea\\
$^{42}$ Gauhati University, Department of Physics, Guwahati, India\\
$^{43}$ Helmholtz-Institut f\"{u}r Strahlen- und Kernphysik, Rheinische Friedrich-Wilhelms-Universit\"{a}t Bonn, Bonn, Germany\\
$^{44}$ Helsinki Institute of Physics (HIP), Helsinki, Finland\\
$^{45}$ High Energy Physics Group,  Universidad Aut\'{o}noma de Puebla, Puebla, Mexico\\
$^{46}$ Horia Hulubei National Institute of Physics and Nuclear Engineering, Bucharest, Romania\\
$^{47}$ HUN-REN Wigner Research Centre for Physics, Budapest, Hungary\\
$^{48}$ Indian Institute of Technology Bombay (IIT), Mumbai, India\\
$^{49}$ Indian Institute of Technology Indore, Indore, India\\
$^{50}$ INFN, Laboratori Nazionali di Frascati, Frascati, Italy\\
$^{51}$ INFN, Sezione di Bari, Bari, Italy\\
$^{52}$ INFN, Sezione di Bologna, Bologna, Italy\\
$^{53}$ INFN, Sezione di Cagliari, Cagliari, Italy\\
$^{54}$ INFN, Sezione di Catania, Catania, Italy\\
$^{55}$ INFN, Sezione di Padova, Padova, Italy\\
$^{56}$ INFN, Sezione di Pavia, Pavia, Italy\\
$^{57}$ INFN, Sezione di Torino, Turin, Italy\\
$^{58}$ INFN, Sezione di Trieste, Trieste, Italy\\
$^{59}$ Inha University, Incheon, Republic of Korea\\
$^{60}$ Institute for Gravitational and Subatomic Physics (GRASP), Utrecht University/Nikhef, Utrecht, Netherlands\\
$^{61}$ Institute of Experimental Physics, Slovak Academy of Sciences, Ko\v{s}ice, Slovak Republic\\
$^{62}$ Institute of Physics, Homi Bhabha National Institute, Bhubaneswar, India\\
$^{63}$ Institute of Physics of the Czech Academy of Sciences, Prague, Czech Republic\\
$^{64}$ Institute of Space Science (ISS), Bucharest, Romania\\
$^{65}$ Institut f\"{u}r Kernphysik, Johann Wolfgang Goethe-Universit\"{a}t Frankfurt, Frankfurt, Germany\\
$^{66}$ Instituto de Ciencias Nucleares, Universidad Nacional Aut\'{o}noma de M\'{e}xico, Mexico City, Mexico\\
$^{67}$ Instituto de F\'{i}sica, Universidade Federal do Rio Grande do Sul (UFRGS), Porto Alegre, Brazil\\
$^{68}$ Instituto de F\'{\i}sica, Universidad Nacional Aut\'{o}noma de M\'{e}xico, Mexico City, Mexico\\
$^{69}$ iThemba LABS, National Research Foundation, Somerset West, South Africa\\
$^{70}$ Jeonbuk National University, Jeonju, Republic of Korea\\
$^{71}$ Johann-Wolfgang-Goethe Universit\"{a}t Frankfurt Institut f\"{u}r Informatik, Fachbereich Informatik und Mathematik, Frankfurt, Germany\\
$^{72}$ Korea Institute of Science and Technology Information, Daejeon, Republic of Korea\\
$^{73}$ KTO Karatay University, Konya, Turkey\\
$^{74}$ Laboratoire de Physique Subatomique et de Cosmologie, Universit\'{e} Grenoble-Alpes, CNRS-IN2P3, Grenoble, France\\
$^{75}$ Lawrence Berkeley National Laboratory, Berkeley, California, United States\\
$^{76}$ Lund University Department of Physics, Division of Particle Physics, Lund, Sweden\\
$^{77}$ Nagasaki Institute of Applied Science, Nagasaki, Japan\\
$^{78}$ Nara Women{'}s University (NWU), Nara, Japan\\
$^{79}$ National and Kapodistrian University of Athens, School of Science, Department of Physics , Athens, Greece\\
$^{80}$ National Centre for Nuclear Research, Warsaw, Poland\\
$^{81}$ National Institute of Science Education and Research, Homi Bhabha National Institute, Jatni, India\\
$^{82}$ National Nuclear Research Center, Baku, Azerbaijan\\
$^{83}$ National Research and Innovation Agency - BRIN, Jakarta, Indonesia\\
$^{84}$ Niels Bohr Institute, University of Copenhagen, Copenhagen, Denmark\\
$^{85}$ Nikhef, National institute for subatomic physics, Amsterdam, Netherlands\\
$^{86}$ Nuclear Physics Group, STFC Daresbury Laboratory, Daresbury, United Kingdom\\
$^{87}$ Nuclear Physics Institute of the Czech Academy of Sciences, Husinec-\v{R}e\v{z}, Czech Republic\\
$^{88}$ Oak Ridge National Laboratory, Oak Ridge, Tennessee, United States\\
$^{89}$ Ohio State University, Columbus, Ohio, United States\\
$^{90}$ Physics department, Faculty of science, University of Zagreb, Zagreb, Croatia\\
$^{91}$ Physics Department, Panjab University, Chandigarh, India\\
$^{92}$ Physics Department, University of Jammu, Jammu, India\\
$^{93}$ Physics Program and International Institute for Sustainability with Knotted Chiral Meta Matter (SKCM2), Hiroshima University, Hiroshima, Japan\\
$^{94}$ Physikalisches Institut, Eberhard-Karls-Universit\"{a}t T\"{u}bingen, T\"{u}bingen, Germany\\
$^{95}$ Physikalisches Institut, Ruprecht-Karls-Universit\"{a}t Heidelberg, Heidelberg, Germany\\
$^{96}$ Physik Department, Technische Universit\"{a}t M\"{u}nchen, Munich, Germany\\
$^{97}$ Politecnico di Bari and Sezione INFN, Bari, Italy\\
$^{98}$ Research Division and ExtreMe Matter Institute EMMI, GSI Helmholtzzentrum f\"ur Schwerionenforschung GmbH, Darmstadt, Germany\\
$^{99}$ Saga University, Saga, Japan\\
$^{100}$ Saha Institute of Nuclear Physics, Homi Bhabha National Institute, Kolkata, India\\
$^{101}$ School of Physics and Astronomy, University of Birmingham, Birmingham, United Kingdom\\
$^{102}$ Secci\'{o}n F\'{\i}sica, Departamento de Ciencias, Pontificia Universidad Cat\'{o}lica del Per\'{u}, Lima, Peru\\
$^{103}$ Stefan Meyer Institut f\"{u}r Subatomare Physik (SMI), Vienna, Austria\\
$^{104}$ SUBATECH, IMT Atlantique, Nantes Universit\'{e}, CNRS-IN2P3, Nantes, France\\
$^{105}$ Sungkyunkwan University, Suwon City, Republic of Korea\\
$^{106}$ Suranaree University of Technology, Nakhon Ratchasima, Thailand\\
$^{107}$ Technical University of Ko\v{s}ice, Ko\v{s}ice, Slovak Republic\\
$^{108}$ The Henryk Niewodniczanski Institute of Nuclear Physics, Polish Academy of Sciences, Cracow, Poland\\
$^{109}$ The University of Texas at Austin, Austin, Texas, United States\\
$^{110}$ Universidad Aut\'{o}noma de Sinaloa, Culiac\'{a}n, Mexico\\
$^{111}$ Universidade de S\~{a}o Paulo (USP), S\~{a}o Paulo, Brazil\\
$^{112}$ Universidade Estadual de Campinas (UNICAMP), Campinas, Brazil\\
$^{113}$ Universidade Federal do ABC, Santo Andre, Brazil\\
$^{114}$ Universitatea Nationala de Stiinta si Tehnologie Politehnica Bucuresti, Bucharest, Romania\\
$^{115}$ University of Cape Town, Cape Town, South Africa\\
$^{116}$ University of Derby, Derby, United Kingdom\\
$^{117}$ University of Houston, Houston, Texas, United States\\
$^{118}$ University of Jyv\"{a}skyl\"{a}, Jyv\"{a}skyl\"{a}, Finland\\
$^{119}$ University of Kansas, Lawrence, Kansas, United States\\
$^{120}$ University of Liverpool, Liverpool, United Kingdom\\
$^{121}$ University of Science and Technology of China, Hefei, China\\
$^{122}$ University of South-Eastern Norway, Kongsberg, Norway\\
$^{123}$ University of Tennessee, Knoxville, Tennessee, United States\\
$^{124}$ University of the Witwatersrand, Johannesburg, South Africa\\
$^{125}$ University of Tokyo, Tokyo, Japan\\
$^{126}$ University of Tsukuba, Tsukuba, Japan\\
$^{127}$ Universit\"{a}t M\"{u}nster, Institut f\"{u}r Kernphysik, M\"{u}nster, Germany\\
$^{128}$ Universit\'{e} Clermont Auvergne, CNRS/IN2P3, LPC, Clermont-Ferrand, France\\
$^{129}$ Universit\'{e} de Lyon, CNRS/IN2P3, Institut de Physique des 2 Infinis de Lyon, Lyon, France\\
$^{130}$ Universit\'{e} de Strasbourg, CNRS, IPHC UMR 7178, F-67000 Strasbourg, France, Strasbourg, France\\
$^{131}$ Universit\'{e} Paris-Saclay, Centre d'Etudes de Saclay (CEA), IRFU, D\'{e}partment de Physique Nucl\'{e}aire (DPhN), Saclay, France\\
$^{132}$ Universit\'{e}  Paris-Saclay, CNRS/IN2P3, IJCLab, Orsay, France\\
$^{133}$ Universit\`{a} degli Studi di Foggia, Foggia, Italy\\
$^{134}$ Universit\`{a} del Piemonte Orientale, Vercelli, Italy\\
$^{135}$ Universit\`{a} di Brescia, Brescia, Italy\\
$^{136}$ Variable Energy Cyclotron Centre, Homi Bhabha National Institute, Kolkata, India\\
$^{137}$ Warsaw University of Technology, Warsaw, Poland\\
$^{138}$ Wayne State University, Detroit, Michigan, United States\\
$^{139}$ Yale University, New Haven, Connecticut, United States\\
$^{140}$ Yonsei University, Seoul, Republic of Korea\\
$^{141}$  Zentrum  f\"{u}r Technologie und Transfer (ZTT), Worms, Germany\\
$^{142}$ Affiliated with an institute covered by a cooperation agreement with CERN\\
$^{143}$ Affiliated with an international laboratory covered by a cooperation agreement with CERN.\\

\end{flushleft} 
  
\end{document}